\documentclass{lmcs}

\usepackage{booktabs}
\usepackage{array}
\usepackage{url}
\usepackage{xcolor}
\usepackage{tikz}
\usetikzlibrary{arrows.meta,positioning}
\usepackage{longtable}

\begin{document}

\title[Formal Grammars in BPM: A Systematic Literature Review]{%
  Formal Grammars in Business Process Management:\\
  A Systematic Literature Review}

\author[M.M.~Zekeng Ndadji]{%
  Milliam Maxime Zekeng Ndadji\lmcsorcid{0000-0002-0417-5591}}

\address{Department of Mathematics and Computer Science,
  University of Dschang, PO Box~67, Dschang, Cameroon}
\email{ndadji.maxime@univ-dschang.org}


\begin{abstract}
Business Process Management~(BPM) is concerned with the systematic design,
execution, monitoring, and improvement of business processes.
Formal grammars have emerged
as a particularly fruitful formalism for BPM, offering generative, declarative,
and analytical capabilities that are uniquely well-suited to process-oriented concerns.
This paper presents a systematic literature review of 34 primary studies at the
intersection of formal grammars and BPM.
We identify seven research streams: (i)~process grammars for organizational process
design; (ii)~process modeling languages evaluated as grammars under the Bunge-Wand-Weber
ontological framework; (iii)~production-rule grammars for process structural specification
and variant management; (iv)~attribute grammars for the declarative specification and
distributed execution of workflows; (v)~graph grammars for the transformation, generation,
and semantic analysis of process models; (vi)~grammatical inference for process mining
and discovery; and (vii)~process algebras as
grammar-like compositional frameworks for behavioral specification and verification.
For each stream, we synthesize contributions, formalisms employed, and limitations.
The review reveals that formal grammars have influenced BPM across every lifecycle
phase (from organizational design to formal verification and data-driven discovery) yet the seven streams have developed largely in parallel, without cross-stream synthesis. We identify five corpus-grounded open challenges and argue that a deeper, unified exploitation of grammatical theory holds significant promise for advancing the state of the art in BPM.
\end{abstract}

\maketitle

\section{Introduction}
\label{sec:introduction}

Business Process Management~(BPM) is a well-established discipline concerned with
the analysis, design, execution, monitoring, and continuous improvement of business
processes~\cite{vanderaalst2004bpm,dumas2018fundamental}.
Over the past four decades, BPM has evolved from ad-hoc workflow automation into a
principled field of research and practice, providing methods and tools for the systematic
management of the operational core of organizations.
The fundamental challenge of BPM is to bridge the gap between the informal knowledge
that domain experts have about how work is done and the formal specifications required
to automate, verify, and optimize that work.
Formal grammars, introduced in theoretical computer science by Chomsky~\cite{chomsky1959certain}
and subsequently enriched by a rich theory of string, tree, and graph rewriting systems, constitute one of the most powerful and flexible mathematical tools available for describing structured entities. A grammar specifies a potentially infinite set of admissible structures (a language) by means of a finite set of rewrite rules applied to an initial symbol. This generative character makes grammars particularly natural for describing processes: just as a grammar generates the sentences of a language, a process grammar can generate all the admissible execution traces of a business process.

The connection between grammars and processes is not merely metaphorical.
At the most basic level, the sequential, parallel, and iterative patterns that
characterize process control flow correspond closely to the sequential composition,
shuffle, and Kleene closure operators that underlie formal language theory.
At a deeper level, attributed grammars provide a natural mechanism for integrating
data-flow and role-assignment alongside structural control-flow, addressing one of
the perennial weaknesses of traditional workflow formalisms.
Graph grammars, on the other hand, are ideally suited to the graphical nature of
process modeling notations such as Business Process Model and Notation (BPMN)~\cite{omg2013bpmn202} and Yet Another Workflow Language (YAWL)~\cite{van2005yawl}, enabling rigorous specification of both syntax and semantics. Finally, grammatical inference techniques offer a principled bridge between empirical process data (event logs) and declarative process models, opening the door to grammar-based process mining.

Despite this natural and multifaceted connection, no comprehensive survey of the
role of formal grammars in BPM exists to date, to the best of our knowledge.
Partial surveys of BPM formalism do exist~\cite{borger2012approaches}, but they
treat grammars as one tool among many rather than as a central object of study,
and systematic grammatical comparisons are absent.
The present paper fills this gap by providing a systematic review of the literature
at the intersection of formal grammars and BPM, spanning works from the mid-1990s
to the mid-2020s.
Our review identifies seven distinct research streams and synthesizes their
contributions, formalisms, and open challenges.
The specific contributions of this paper are as follows.
First, we provide, to the best of our knowledge, the first comprehensive taxonomy of
grammatical approaches in BPM, identifying seven streams of research.
Second, we synthesize the key findings of each stream, highlighting the specific
grammatical formalisms employed and the BPM lifecycle phases addressed.
Third, we identify cross-cutting themes, research gaps, and directions for future work.

The remainder of the paper is organized as follows.
Section~\ref{sec:background} provides the necessary background on BPM and formal
grammars.
Section~\ref{sec:methodology} describes the review methodology.
Sections~\ref{sec:stream1}--\ref{sec:stream7} present the seven identified research streams.
Section~\ref{sec:discussion} provides a cross-cutting discussion and identifies open
challenges.
Section~\ref{sec:conclusion} concludes the paper.

\section{Background}
\label{sec:background}

\subsection{Business Process Management}

A business process is a set of activities that, together, achieve a specific
organizational goal~\cite{dumas2018fundamental}.
The Workflow Management Coalition defines a workflow as the automation of a
business process, in whole or part, during which documents, information, or tasks
are passed from one participant to another for action, according to a set of
procedural rules~\cite{coalition1999workflow}.
BPM provides the methods, tools, and systems to support the design, enactment,
management, and analysis of such processes.
The BPM lifecycle typically comprises five phases: \emph{design and analysis},
\emph{configuration}, \emph{enactment}, \emph{evaluation}, and
\emph{adaptation}~\cite{dumas2018fundamental}.
In the design and analysis phase, processes are discovered, documented, and modeled using workflow languages.
In the configuration phase, the resulting models are deployed onto a
Workflow Management System~(WfMS).
In the enactment phase, process instances are executed and monitored.
In the evaluation phase, performance data is collected and analysed (sub-activities
of this phase include formal \emph{verification} of process correctness and
\emph{monitoring} of running instances) and in the adaptation phase, the process is
improved accordingly; \emph{process discovery} (automated model extraction from
event logs) bridges adaptation back into design.
These sub-activities are treated as distinct BPM concerns in the inclusion criteria
and research questions of this review.

Numerous workflow languages have been proposed to support these phases.
Petri nets, notably in their workflow net~(WF-net) variant~\cite{van1996structural},
provide a formal, graph-based foundation for process modeling with well-understood
semantics.
YAWL~\cite{van2005yawl}, designed to address the expressiveness requirements captured
by the workflow patterns catalogue~\cite{van2003workflow}, extends WF-nets with
additional control-flow constructs.
Event-driven Process Chains~(EPCs)~\cite{scheer2005process} support process modeling
in enterprise information systems.
The BPMN~\cite{omg2013bpmn202}, currently
the de-facto standard for business process modeling, provides an extensive set of
graphical constructs for describing processes in a notation accessible to both
technical and non-technical stakeholders.
A key weakness of many of these languages, however, is the informality of their
semantics~\cite{dijkman2008semantics,borger2012approaches}, which
impedes formal analysis and verification.
The artifact-centric approach~\cite{nigam2003business,bhattacharya2007artifact}
represents an alternative paradigm in which processes are driven by the data objects
(artifacts) they manipulate, rather than by a fixed control flow.
This approach is particularly relevant for our review, as it creates a natural
opportunity for attribute grammar-based specification, where attributes on grammar
symbols naturally capture the data fields of artifacts.

\subsection{Formal Grammars: A Brief Overview}
\label{subsec:grammars}

A formal grammar $G = (V, \Sigma, R, S)$ consists of a finite set $V$ of variables
(non-terminals), a finite set $\Sigma$ of terminal symbols (disjoint from $V$),
a finite set $R$ of production rules, and a distinguished start symbol $S \in V$.
The language $L(G)$ generated by $G$ is the set of all strings over $\Sigma$
derivable from $S$ by applying rules in $R$.
The Chomsky hierarchy~\cite{chomsky1959certain,hopcroft2007introduction} classifies
string grammars into four types by the form of their production rules.
\emph{Regular grammars}~(Type~3) generate regular languages, recognized by finite
automata; they capture sequential iteration but not nesting.
\emph{Context-free grammars}~(CFGs, Type~2) allow rules of the form $A \to \alpha$,
where $A$ is a non-terminal and $\alpha$ is any string over $V \cup \Sigma$; their
languages are recognized by pushdown automata and naturally capture hierarchical,
nested structure.
\emph{Context-sensitive grammars}~(Type~1) are equivalently characterized by
two standard formulations~\cite{hopcroft2007introduction}.
In the restricted form, each rule has the shape
$\alpha A \beta \to \alpha \gamma \beta$
where $A$ is a non-terminal, $\alpha, \beta \in (V \cup \Sigma)^{*}$, and
$\gamma \in (V \cup \Sigma)^{+}$, so $A$ is rewritten only in the surrounding context
$(\alpha, \beta)$.
In the equivalent non-contracting form, any rule $u \to v$ with $|u| \leq |v|$
is permitted (with the single exception $S \to \varepsilon$, allowed only when $S$
does not appear on the right-hand side of any rule).
Both forms generate the same class of languages, which lies strictly between the
context-free and recursively enumerable languages.
\emph{Unrestricted grammars}~(Type~0) have no restrictions on their rules and
generate recursively enumerable languages.

\emph{Attribute grammars}, introduced by Knuth~\cite{knuth1968semantics}, extend
CFGs with semantic attributes associated with each grammar symbol.
Each production rule is augmented with semantic equations that specify how
attribute values propagate: synthesized attributes flow bottom-up from
children to parents, while inherited attributes flow top-down.
Attribute grammars provide a principled mechanism for specifying the static semantics
of formal languages and, by extension, for integrating data computation with
structural specification in process models.

\emph{Graph grammars}~\cite{ehrig1999handbook} generalize string grammars to
graph-structured objects.
Rather than rewriting strings of symbols, graph grammars rewrite graphs by replacing
sub-graphs according to production rules.
Major formalisms include hyperedge replacement grammars~(HRGs), in which
rewriting is performed on hypergraphs, node-label-controlled~(NLC) grammars,
and algebraic approaches based on the single-pushout~(SPO) or
double-pushout~(DPO) constructions.
Graph grammars are particularly well-suited to BPM because process models are
inherently graph-structured: nodes represent activities or events, and edges
represent control or data flow.

\emph{Grammatical inference}~(also called \emph{grammar induction}) is the problem
of learning a grammar from a sample of positive (and possibly negative) examples.
Classical results establish the limits of identification in the limit from positive
data: Gold's theorem~\cite{hopcroft2007introduction} shows that any class of languages
that contains all finite languages and at least one infinite language (including the
regular languages and the context-free languages) cannot in general be
identified in the limit from positive examples alone.
For regular languages, however, efficient algorithms such as RPNI~\cite{oncina1992rpni}
succeed when the target language satisfies structural constraints derivable from
the sample; for context-free languages, identification requires either negative
examples, structural constraints (e.g., bracket annotations), or probabilistic relaxations.
In the context of process mining, event logs constitute a finite corpus of positive
execution traces, making grammatical inference a natural computational framework for
process model discovery.

\section{Review Methodology}
\label{sec:methodology}

This review follows the systematic literature review~(SLR) protocol described
by Kitchenham and Charters~\cite{kitchenham2007guidelines} for software
engineering research, adapted to the BPM and formal language theory context.
The protocol comprises four phases: database search, screening and selection,
classification, and synthesis.

\subsection{Research Questions}

This review is organized around four research questions derived from the gap
identified in Section~\ref{sec:introduction}:

\begin{description}
\item[RQ1] What types of formal grammar formalisms (string, attribute, graph,
  inferential, or process-algebraic) have been applied in BPM research, and how
  do they relate to the Chomsky hierarchy?
\item[RQ2] For which phases of the BPM lifecycle (design and analysis,
  configuration, enactment, evaluation, or adaptation---including verification,
  process discovery, and monitoring as analytical sub-activities) are formal
  grammar formalisms employed?
\item[RQ3] What are the key contributions and limitations of each identified
  research stream, and how do the streams compare with respect to formalism
  expressiveness and BPM coverage?
\item[RQ4] What are the principal open research challenges at the intersection of
  formal grammar theory and BPM, and what tools or methods are available to
  address them?
\end{description}

These questions determine the search strategy~(RQ1--RQ2), the inclusion
criteria~(RQ1--RQ2), the synthesis structure~(RQ3), and the discussion
section~(RQ4).

\subsection{Search Strategy}

Six electronic databases were queried in February 2026:
ACM Digital Library, IEEE Xplore, SpringerLink, ScienceDirect, Scopus, and
Google Scholar.
The following Boolean search strings were used systematically in each database:
\begin{itemize}
\item \texttt{"formal grammar" AND "business process"};
\item \texttt{"process grammar"};
\item \texttt{"graph grammar" AND ("workflow" OR "business process")};
\item \texttt{"attribute grammar" AND "workflow"};
\item \texttt{"grammatical inference" AND "process mining"};
\item \texttt{"process modeling grammar"};
\item \texttt{"context-free grammar" AND "business process"};
\item \texttt{"hyperedge replacement grammar" AND "process"}.
\end{itemize}
No publication-date restriction was applied.
The search returned 526~records in total
(ACM~DL: 74; IEEE~Xplore: 68; SpringerLink: 112;
ScienceDirect: 93; Scopus: 101; Google~Scholar: 78).

\subsection{Inclusion and Exclusion Criteria}

A paper was included if it satisfied both of the following criteria:
(IC1)~it proposed, applied, analysed, or formally grounded at least one
grammar-based formalism, whether a technically operative grammar (string,
attribute, or graph grammar, a grammatical inference algorithm, or a
process-algebraic formalism whose term constructors constitute an inductive
grammar over process expressions) or a conceptual grammar framework explicitly
constructed in terms of a lexicon, rewrite rules, and constraints in direct
analogy with a recognized class of formal grammars;
and (IC2)~its principal contribution addressed at least one phase of the
BPM lifecycle, including the sub-activities of verification, process discovery,
and monitoring as defined in Section~\ref{sec:background}.

A paper was excluded if any of the following held:
(EC1)~grammars were used only peripherally, e.g., for parsing configuration
files, XML schemas, or general programming-language syntax, or as an
implementation artifact of a contribution whose primary object of study was
not the grammar formalism itself (for example, process discovery algorithms
that internally produce process trees without treating the tree grammar as the
central formalism were excluded under this criterion);
(EC2)~the BPM content was limited to a passing motivating example without a
substantive technical contribution; or
(EC3)~the full text was inaccessible through institutional subscriptions or
open-access repositories.

\subsection{Screening and Selection}

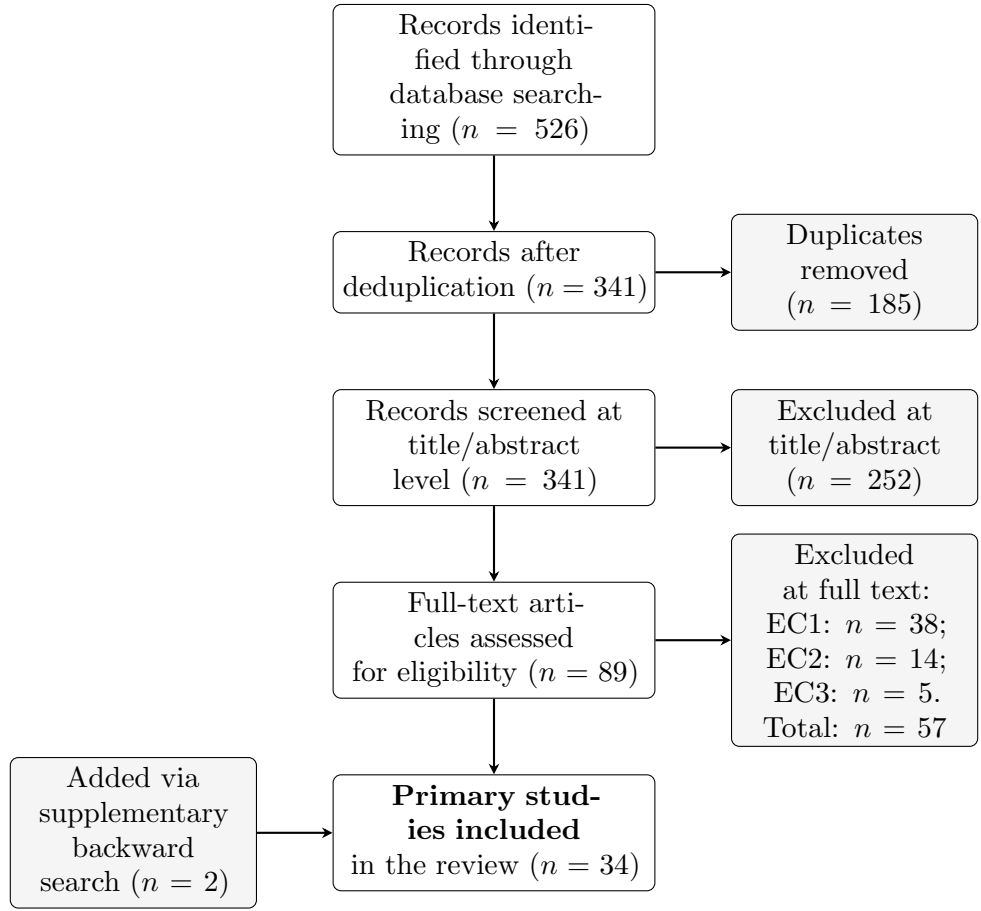
\begin{figure}[ht]
\centering
\begin{tikzpicture}[
  mainbox/.style={draw, rectangle, rounded corners=3pt, text width=4cm,
                  align=center, minimum height=1.05cm},
  sidebox/.style={draw, rectangle, rounded corners=3pt, text width=3cm,
                  align=center, minimum height=1.0cm, fill=gray!8},
  arr/.style={-{Stealth[length=4pt,width=4pt]}, thick}
]
\node[mainbox] (A)
  {Records identified through\\database searching ($n=526$)};
\node[mainbox, below=1cm of A] (B)
  {Records after\\deduplication ($n=341$)};
\node[mainbox, below=1cm of B] (C)
  {Records screened at\\title/abstract level ($n=341$)};
\node[mainbox, below=1cm of C] (D)
  {Full-text articles assessed\\for eligibility ($n=89$)};
\node[mainbox, below=1cm of D] (E)
  {\textbf{Primary studies included}\\in the review ($n=34$)};

\node[sidebox, right=1cm of B] (R1)
  {Duplicates removed\\($n=185$)};
\node[sidebox, right=1cm of C] (R2)
  {Excluded at title/abstract\\($n=252$)};
\node[sidebox, right=1cm of D] (R3)
  {Excluded at full text:\\EC1: $n=38$;\ EC2: $n=14$;\\EC3: $n=5$.\ \ Total: $n=57$};
\node[sidebox, left=1cm of E] (S1)
  {Added via supplementary\\backward search ($n=2$)};

\draw[arr] (A) -- (B);
\draw[arr] (B) -- (C);
\draw[arr] (C) -- (D);
\draw[arr] (D) -- (E);
\draw[arr] (B.east) -- (R1.west);
\draw[arr] (C.east) -- (R2.west);
\draw[arr] (D.east) -- (R3.west);
\draw[arr] (S1.east) -- (E.west);
\end{tikzpicture}
\caption{PRISMA-style flow diagram of the literature search and selection process.}
\label{fig:prisma}
\end{figure}

After removing 185 duplicates, 341 unique records remained.
Title and abstract screening eliminated 252 records that clearly did not satisfy
IC1 and IC2, leaving 89 candidates for full-text assessment.
Research monographs were included when they constituted original, self-contained
research contributions rather than textbooks or edited collections; study~\#8
(Recker~\cite{recker2011book}) satisfies this criterion, combining a novel
ontological analysis of BPMN with primary qualitative and quantitative data
not published elsewhere.
Full-text reading resulted in the further exclusion of 57 papers
(38 for EC1; 14 for EC2; 5 for EC3), yielding 32 primary studies
from the main search.
A supplementary backward citation search, conducted after the initial screening
and focused on process algebra approaches to BPM (additional query strings:
\texttt{"process algebra" AND "workflow"} and
\texttt{("CSP" OR "pi-calculus") AND "business process"}),
identified two further primary studies on $\pi$-calculus
semantics~\cite{puhlmann2005pi,lucchi2007pi} that satisfied all inclusion
criteria but had not surfaced in the main database search.
These two studies were added, bringing the final corpus to 34 primary
studies, as reflected in the PRISMA diagram~(Figure~\ref{fig:prisma}).
Figure~\ref{fig:prisma} summarises the selection process as a PRISMA-style
flow diagram~\cite{moher2009preferred}.

A formal quality assessment of primary studies---rating each study on a
checklist of methodological rigour criteria (precision of formal definitions,
presence of proofs or empirical evaluation, explicit limitation
acknowledgement)---was not conducted for this review.
This is a known limitation relative to the Kitchenham~\&~Charters
protocol~\cite{kitchenham2007guidelines}, which recommends quality
scoring to allow differential weighting of evidence.
The decision was pragmatic: the 34 primary studies span three decades
and seven research streams, each associated with distinct methodological
traditions (formal proofs, empirical surveys, case studies, prototype
implementations), and this heterogeneity makes a single quality scale
difficult to operationalize without introducing additional reviewer bias.
Each stream section instead applies a consistent critical template
(contributions, limitations, comparative analysis) that provides a
qualitative proxy for study quality assessment.

Additionally, 12 foundational reference works (standard textbooks,
language specifications, and seminal papers on BPM theory or grammar theory
that were not primary contributions to the grammar/BPM intersection but were
essential as background) were retained as supporting references.
These are: \cite{chomsky1959certain, knuth1968semantics, ehrig1999handbook,
kitchenham2007guidelines, wand1993anchoring, van1996structural, van2003workflow,
van2005yawl, omg2013bpmn202, dumas2018fundamental, coalition1999workflow,
scheer2005process}.

\subsection{Classification}
\label{subsec:classification}

The 34 primary studies were classified into thematic streams using a card-sorting procedure: each study was assigned a descriptive label
based on~(a) the grammar formalism employed and~(b) the principal BPM concern
addressed.
Labels were then grouped inductively into streams.
Streams were distinguished on the basis of formalism family and primary BPM
concern rather than cluster size: a cohesive group of studies was classified
as an independent stream when its core formalism (whether a Chomsky-hierarchy
string class, a graph rewriting system, or a process algebra) was formally and
conceptually distinct from those of all other clusters, regardless of the number
of studies it contained.
Stream~VII~(2 studies) satisfies this criterion because process algebras
constitute a formally separate grammatical paradigm from the string and graph
formalisms of Streams~I--VI; the small size of this cluster is explicitly
reported as a limitation in Challenge~C3 (Section~\ref{sec:discussion}).
Because this classification was performed by a single reviewer, the absence
of a second rater and a formal inter-rater reliability measure~(e.g.,
Cohen's~$\kappa$) is a known limitation of the present review.
To partially mitigate this, the classification was cross-validated against
the self-descriptions in each study's abstract and keywords, and was checked
for coherence against the major BPM and formal-methods survey literature.

Seven streams emerged from this procedure:
(I)~process grammars for organizational design;
(II)~process modeling languages analyzed as grammars;
(III)~string grammars for process specification and variants;
(IV)~attribute grammars for workflow specification and execution;
(V)~graph grammars for process model transformation and analysis;
(VI)~grammatical inference for process mining;
and (VII)~process algebras as grammatical frameworks for behavioral specification.
One study, Datta~\cite{datta1998automating}, was assigned to two streams~(I and~VI)
because it makes independent contributions to both: it applies a grammatical metaphor
to process design~(Stream~I) and introduces grammar-discovery algorithms for
automated process model extraction~(Stream~VI).
It is discussed in both sections, with each discussion focusing on the contribution
relevant to that stream.
This dual assignment does not affect the corpus count: the study is counted once
(as primary study~\#6) in Table~\ref{tab:complete}.
Table~\ref{tab:complete} provides a complete enumeration of all 34 primary
studies with their venue type, stream assignment, grammar formalism,
and BPM lifecycle phase.
Table~\ref{tab:taxonomy} gives a compact classification overview.
Table~\ref{tab:temporal} presents the temporal distribution of the 34 primary
studies by stream and five-year period.
Three phases of activity are discernible: a formative phase~(1994--2004, $n=5$)
dominated by Stream~I process grammars; a growth phase~(2005--2014, $n=15$) in
which Streams~II, V, and~VII emerged; and a recent diversification phase~(2015--2024,
$n=14$) during which Streams~III, IV, and~VI became dominant.
Streams~IV and~VI together account for six of the seven studies published in
2020--2024, suggesting that, in the most recent period covered by the corpus,
attributed-grammar workflow specification and stochastic grammatical inference may
constitute the most dynamically expanding frontiers.

\small
\begin{longtable}{@{}rp{5.85cm}cccp{1.9cm}p{1.4cm}@{}}
\caption{Complete list of 34 primary studies included in the review.
Venue type: J~=~journal; C~=~conference/workshop;
B~=~book; R~=~technical report.
Grammar: PG~=~process grammar; MG~=~modeling grammar (ontological);
CFG~=~context-free grammar; CSG~=~context-sensitive grammar;
AG~=~attribute grammar; GAG~=~guarded AG; TGG~=~triple graph grammar;
HRG~=~hyperedge replacement grammar; EGGG~=~extended edge-based graph grammar;
HOT~=~higher-order graph transformation; PGG~=~process graph grammar;
PCFG~=~probabilistic CFG; GI~=~grammatical inference.
BPM phase: De~=~design; Mo~=~modeling; Ex~=~execution; Ve~=~verification;
Di~=~discovery; Mon~=~monitoring; Tr~=~model transformation.}
\label{tab:complete}\\
\toprule
\textbf{\#} & \textbf{Authors} & \textbf{Year} & \textbf{Type} & \textbf{Stream} & \textbf{Formalism} & \textbf{BPM Phase} \\
\midrule
\endfirsthead
\multicolumn{7}{l}{\small\itshape Table~\ref{tab:complete} continued.}\\
\toprule
\# & Authors & Year & Type & Stream & Formalism & BPM Phase \\
\midrule
\endhead
\midrule
\multicolumn{7}{r}{\small\itshape Continued on next page.}\\
\endfoot
\bottomrule
\endlastfoot
 1 & Pentland~\cite{pentland1994process}                   & 1994 & R & I      & PG         & De \\
 2 & Pentland~\cite{pentland1995grammatical}               & 1995 & J & I      & PG         & De \\
 3 & Pentland \& Rueter~\cite{pentland1994routines}        & 1994 & J & I      & PG         & De \\
 4 & Lee \& Pentland~\cite{lee2003exploring}               & 2003 & R & I      & PG         & De \\
 5 & Lee et~al.~\cite{lee2008process}                      & 2008 & J & I      & PG         & De \\
 6 & Datta~\cite{datta1998automating}                      & 1998 & J & I, VI  & PG / GI    & De, Di \\
\midrule
 7 & Rosemann et~al.~\cite{rosemann2006evolution}            & 2006 & C & II     & MG         & Mo \\
 8 & Recker~\cite{recker2011book}                          & 2011 & B & II     & MG         & Mo \\
 9 & Recker et~al.~\cite{recker2010ontological}            & 2010 & J & II     & MG         & Mo \\
10 & Recker \& Rosemann~\cite{recker2010measurement}       & 2010 & J & II     & MG         & Mo \\
11 & Recker et~al.~\cite{recker2011misq}                   & 2011 & J & II     & MG         & Mo \\
12 & Recker~\cite{recker2012modeling}                      & 2012 & J & II     & MG         & Mo \\
13 & Recker~\cite{recker2013gateway}                       & 2013 & J & II     & MG         & Mo \\
\midrule
14 & Ayari et~al.~\cite{ayari2019grammar}                  & 2019 & J & III    & CFG        & Mo, Ve \\
15 & Sarno et~al.~\cite{sarno2015context}                  & 2015 & J & III    & CSG        & Mo \\
16 & Montero et~al.~\cite{montero2008feature}              & 2008 & C & III    & Feature    & De \\
17 & Van~der~Aa et~al.~\cite{vanderaa2019extracting}       & 2019 & C & III    & NLP        & Mo \\
\midrule
18 & Badouel et~al.~\cite{badouel2014grammatical}          & 2014 & R & IV     & GAG        & Ex \\
19 & Badouel et~al.~\cite{badouel2015active}               & 2015 & J & IV     & GAG        & Ex \\
20 & Zekeng Ndadji et~al.~\cite{ndadji2020language}        & 2020 & J & IV     & AG  & Mo, Ex \\
21 & Zekeng Ndadji et~al.~\cite{zekeng2021projection}      & 2021 & J & IV     & AG  & Ex \\
22 & Zekeng Ndadji et~al.~\cite{ndadji2023non}             & 2023 & J & IV     & AG  & Mo \\
23 & Tchoupe Tchendji et~al.~\cite{tchendji2024grammatical}& 2024 & J & IV     & AG  & Mo, Ex \\
\midrule
24 & Lohmann et~al.~\cite{lohmann2007tgg}                  & 2007 & J & V      & TGG        & Tr \\
25 & Mazanek \& Hanus~\cite{mazanek2011bpmn}               & 2011 & J & V      & HRG        & Tr \\
26 & Shi et~al.~\cite{shi2016bpmn}                         & 2016 & J & V      & EGGG       & Tr, Ve \\
27 & Kr\"{a}uter et~al.~\cite{krauter2024higher}           & 2024 & J & V      & HOT        & Ve \\
28 & Kataeva \& Kalenkova~\cite{kataeva2019graph}           & 2019 & C & V      & PGG        & Di \\
\midrule
29 & Bergenthum et~al.~\cite{bergenthum2007process}        & 2007 & C & VI     & Region/GI  & Di \\
30 & Breuker et~al.~\cite{breuker2016comprehensible}       & 2016 & J & VI     & Prob.~GI   & Di, Mon \\
31 & Watanabe et~al.~\cite{watanabe2023grammar}             & 2023 & J & VI     & PCFG       & Mon \\
32 & Alkhammash et~al.~\cite{alkhammash2024stochastic}      & 2024 & C & VI     & GI (ALERGIA) & Di \\
\midrule
33 & Puhlmann \& Weske~\cite{puhlmann2005pi}               & 2005 & C & VII    & $\pi$-calc.  & Mo, Ve \\
34 & Lucchi \& Mazzara~\cite{lucchi2007pi}                 & 2007 & J & VII    & $\pi$-calc.  & Ve \\
\end{longtable}

\begin{table}[ht]
\caption{Classification overview of the seven research streams.}
\label{tab:taxonomy}
\centering\small
\begin{tabular}{@{}lp{3.6cm}p{7.3cm}p{1.4cm}@{}}
\toprule
\textbf{Stream} & \textbf{Grammar type} & \textbf{Representative works} & \textbf{BPM phase} \\
\midrule
I   & Process grammar    & \cite{pentland1995grammatical,lee2008process,datta1998automating}     & De, Di \\
II  & Modeling grammar   & \cite{rosemann2006evolution,recker2011misq,recker2013gateway}           & Mo \\
III & CFG / CSG          & \cite{ayari2019grammar,sarno2015context,vanderaa2019extracting}       & Mo, De \\
IV  & Attribute grammar  & \cite{badouel2014grammatical,ndadji2020language,tchendji2024grammatical} & Mo, Ex \\
V   & Graph grammar      & \cite{lohmann2007tgg,mazanek2011bpmn,krauter2024higher}               & Tr, Ve \\
VI  & Gram.\ inference   & \cite{bergenthum2007process,breuker2016comprehensible,alkhammash2024stochastic} & Di, Mon \\
VII & Process algebra    & \cite{puhlmann2005pi,lucchi2007pi}                    & Mo, Ve \\
\bottomrule
\end{tabular}
\end{table}

\begin{table}[ht]
	\caption{Temporal distribution of the 34 primary studies by research stream and
		five-year period. Study~\#6 (Datta~\cite{datta1998automating}), assigned to both
		Streams~I and~VI, is counted under Stream~I in this table.}
	\label{tab:temporal}
	\centering
	\begin{tabular}{@{}lccccccc|c@{}}
		\toprule
		\textbf{Period} & \textbf{I} & \textbf{II} & \textbf{III} & \textbf{IV} &
		\textbf{V} & \textbf{VI} & \textbf{VII} & \textbf{Total} \\
		\midrule
		1990--1994 & 2 & 0 & 0 & 0 & 0 & 0 & 0 & 2 \\
		1995--1999 & 2 & 0 & 0 & 0 & 0 & 0 & 0 & 2 \\
		2000--2004 & 1 & 0 & 0 & 0 & 0 & 0 & 0 & 1 \\
		2005--2009 & 1 & 1 & 1 & 0 & 1 & 1 & 2 & 7 \\
		2010--2014 & 0 & 6 & 0 & 1 & 1 & 0 & 0 & 8 \\
		2015--2019 & 0 & 0 & 3 & 1 & 2 & 1 & 0 & 7 \\
		2020--2024 & 0 & 0 & 0 & 4 & 1 & 2 & 0 & 7 \\
		\midrule
		\textbf{Total} & 6 & 7 & 4 & 6 & 5 & 4 & 2 & \textbf{34} \\
		\bottomrule
	\end{tabular}
\end{table}

\subsection{Synthesis}

In the following sections (\ref{sec:stream1}--\ref{sec:stream7}), we describe for each stream:
(a)~the motivating problem;
(b)~the specific grammar formalism(s) employed;
(c)~the principal contributions and limitations of the surveyed works;
and~(d)~a within-stream comparison of approaches where multiple works exist.

\section{Stream~I: Process Grammars for Organizational Process Design}
\label{sec:stream1}

\subsection{The Grammatical Metaphor for Organizational Processes}

The systematic application of formal grammar concepts to organizational processes was early explored by Pentland~\cite{pentland1994process,pentland1995grammatical}.
Drawing on Chomsky's theory of generative grammars and Weick's earlier suggestion
that organizing resembles a grammar, Pentland~\cite{pentland1995grammatical}
developed the grammatical metaphor into a rigorous research framework applicable to
empirical studies of organizational processes.
In Pentland's model, a process grammar consists of three components:
a lexicon of elementary actions (called moves), a set of
rewrite rules specifying how moves can be combined, and a set of
constraints that restrict the space of admissible combinations.
Moves are analogous to words in a natural language: they are the atomic units of
observable behavior.
Syntactic constituents (groups of moves that form meaningful sub-processes) play a
role analogous to phrases in a sentence.
A complete process (an observable sequence of events from start to finish) is
analogous to a grammatical sentence.

This framing offers four key theoretical benefits over alternative approaches to
process analysis.
First, it provides a compact representation of potentially infinite behavioral variety: a finite grammar generates an infinite set of admissible process trajectories.
Second, it captures nesting: just as sentences have hierarchical phrase
structure, processes have hierarchical sub-process structure that cannot be captured
by flat sequential models.
Third, it enables classification: different processes (or different instances of the same process) can be compared in terms of the grammars that generate them.
Fourth, it supports design: a grammar can be used generatively to enumerate the space of feasible process alternatives.
This foundational work also established that organizational routines can be
understood as grammars of action~\cite{pentland1994routines}: the recurring patterns
of organizational behavior arise from a constrained generative process, analogous to
the way sentences of a natural language arise from a grammar.

\subsection{Process Grammar as a Design Tool}

Building on the theoretical foundations of~\cite{pentland1995grammatical},
Lee and Pentland~\cite{lee2003exploring} and subsequently Lee, Wyner, and
Pentland~\cite{lee2008process} developed process grammars into a practical tool
for business process design.

Lee and Pentland~\cite{lee2003exploring} argued that existing approaches to process
redesign (including benchmarking, incremental variation, brainstorming, and first-principles engineering) are limited because they do not provide a systematic representation of the space of feasible alternative processes.
A process grammar, by contrast, makes this space explicit, visible, and
searchable, while remaining useful even when the domain representation is incomplete.
The authors illustrated the approach with a grammar for the sales process, demonstrating
how the grammar can be used to generate and compare alternative process designs
in a principled way.

Lee, Wyner, and Pentland~\cite{lee2008process} subsequently operationalized this
framework in a paper.
They introduced two main artifacts: (1)~a grammar-based method for generating and
managing business process design alternatives, and (2)~a prototype software tool
implementing the method.
The method decomposes a process design problem into the choice of a lexicon
(what activities are available), the definition of grammar rules (how activities can
be combined), and the application of domain constraints (which combinations are
feasible or desirable).
The prototype was demonstrated using a grammar for a sales process, showing how the
tool supports designers in exploring the process design space, evaluating alternatives,
and making informed choices.

\subsection{Grammar-Based Automated Process Discovery}

An early contribution to the automation of process model discovery using
grammar-like techniques was made by Datta~\cite{datta1998automating}, who proposed
probabilistic and algorithmic methods for the automatic discovery of AS-IS business
process models.
Datta's work drew on grammar discovery algorithms to extract process models from
observations of organizational activity, addressing a fundamental practical problem:
organizations undertaking Business Process Re\-engineering~(BPR) often lack
well-documented models of their existing processes, and creating such models manually
is expensive and error-prone.
The proposed algorithms are among the earliest attempts to bridge formal language
theory and the BPM domain in a data-driven manner.

\subsection{Comparative Analysis and Limitations}

The three threads within Stream~I represent distinct orientations that together
span the design--discovery axis of BPM.
Pentland's foundational work~\cite{pentland1994process,pentland1995grammatical} is
descriptive: the grammar serves as a theoretical lens for understanding and comparing observed organizational routines, without prescribing or generating specific
designs.
Lee and Pentland~\cite{lee2003exploring,lee2008process} make the grammar
prescriptive: it becomes a tool for generating and evaluating design
alternatives, operationalized in prototype software.
Datta~\cite{datta1998automating} makes the grammar inductive: the grammar structure is recovered algorithmically from observations of organizational activity,
anticipating the process discovery tradition by nearly a decade.
Several common limitations apply across Stream~I.
The process grammar formalism remains at an informal level relative to the Chomsky
hierarchy: no formal language-class membership results are established, and the
connection to string or tree rewriting is conceptual rather than mathematical.
The lexicon of moves must be defined manually by a domain expert, making the approach
domain-specific and limiting transferability across organizational contexts.
Empirical validation in all three threads relies on small, illustrative case studies;
none of the works provides a large-scale evaluation.
Finally, Stream~I addresses exclusively the design and discovery phases of the BPM
lifecycle; enactment, evaluation, and adaptation lie outside its scope.

\section{Stream~II: Process Modeling Languages as Grammars}
\label{sec:stream2}

\subsection{The Ontological Framework for Grammar Evaluation}

A distinct and highly productive line of research treats process modeling languages
(such as BPMN, EPC, UML Activity Diagrams) as modeling grammars: formal
systems consisting of a set of graphical constructs and rules for combining them
to express relevant aspects of business processes~\cite{recker2010ontological}.
Note that modeling grammar in this stream refers to a representational system in the sense of Wand and Weber (a formal notation judged by which ontological constructs it can express, not by the language it generates) and
should be distinguished from the generative sense of Section~\ref{sec:background}.
This perspective, developed primarily by Recker, Rosemann, Indulska, and Green,
uses the Bunge-Wand-Weber~(BWW) ontological theory~\cite{wand1993anchoring}
as a normative benchmark for evaluating the representational adequacy of modeling
grammars.
The BWW model specifies the ontological constructs that any modeling grammar
purporting to model the real world must be capable of representing.
A grammar is said to suffer from (1) construct deficit if it lacks constructs for modeling certain ontologically relevant phenomena, (2) construct overload if a single construct maps to multiple ontological concepts, (3) construct
redundancy if multiple constructs map to the same concept, or (2) construct excess if a construct maps to no ontological concept at all.
Rosemann, Recker, Indulska, and Green~\cite{rosemann2006evolution} were among the first to apply this framework to a comparative study of process modeling grammars, tracing the evolution of representational capabilities across four decades and several generations of process modeling techniques. Their analysis revealed systematic patterns of improvement but also persistent gaps in the representational power of widely used grammars.

\subsection{Empirical Analysis of BPMN as a Modeling Grammar}

Recker~\cite{recker2011book} provided a landmark comprehensive evaluation of BPMN
as a process modeling grammar, combining ontological analysis with qualitative
interviews and quantitative survey methods.
This monograph established a methodological template for rigorous empirical research
on the usability and expressiveness of process modeling grammars.
Recker, Indulska, Rosemann, and Green~\cite{recker2010ontological} published an
empirical investigation of the ontological deficiencies of BPMN in practice.
Through semi-structured interviews with practitioners, they identified nine
ontological deficiencies in BPMN (including the inadequate capture of business
rules and the insufficient support for process decomposition) and five contextual
factors that influence the use of modeling grammars in practice, such as tool support
and organizational modeling conventions.
Recker and Rosemann~\cite{recker2010measurement} developed and validated a
measurement instrument for studying user acceptance of process modeling grammars,
advancing the empirical research agenda in this area.
Recker, Rosemann, Green, and Indulska~\cite{recker2011misq} subsequently used this
instrument in a large-scale survey of 528 practitioners, demonstrating empirically
that ontological deficiencies perceived by users negatively affect their beliefs
about the usefulness and ease of use of process modeling grammars---a result that
validates the BWW-based theoretical framework.
Further studies examined the role of specific grammatical constructs.
Recker~\cite{recker2012modeling} showed that five features of modeling tools
significantly affect users' beliefs about the grammar they work with, demonstrating
that grammar usage cannot be studied in isolation from the tool environment.
Recker~\cite{recker2013gateway} conducted a controlled experiment examining the
usefulness of Gateway constructs (the constructs in BPMN used to express
control-flow divergence and convergence), finding that Gateway constructs improve
the interpretability of process models through a perceptual discriminability effect,
particularly for complex models.

\subsection{Comparative Analysis and Limitations}

The seven studies in Stream~II form a coherent methodological arc rather than a
collection of independent investigations.
Rosemann et al.~\cite{rosemann2006evolution} established the BWW benchmark and applied
it in a comparative historical analysis of multiple modeling grammars.
Recker~\cite{recker2011book} deepened this into a comprehensive evaluation of a single
grammar (BPMN), integrating ontological analysis with qualitative and quantitative data.
Recker et al.~\cite{recker2010ontological} and Recker and
Rosemann~\cite{recker2010measurement} developed and validated the measurement
infrastructure required for large-scale survey research.
Recker et al.~\cite{recker2011misq} then deployed this infrastructure in a survey of
528~practitioners.
Finally, Recker~\cite{recker2012modeling,recker2013gateway} used controlled experiments
to identify the specific grammatical constructs and tool features that drive
practitioner perception.
This progression reflects three systematic shifts: from multiple grammars to a single
grammar~(BPMN); from descriptive analysis (which deficiencies exist) to explanatory
and predictive claims (why they matter and what constructs drive usability); and from
theoretical comparison to surveys to controlled experiments.
The stream's contribution is thus not only substantive but also methodological,
demonstrating a replicable template for the empirical evaluation of process modeling
grammars. It should be noted that all seven studies in this stream were conducted by the
same research group (Recker, Rosemann, Indulska, Green, and collaborators);
Stream~II therefore reflects the depth and sustained productivity of a single
research program rather than a broad community of independent contributors.
This homogeneity limits the generalizability of the stream's findings but
does not diminish their internal coherence or empirical quality.

\section{Stream~III: Production-Rule Grammars for Process Structural Specification and Variant Management}
\label{sec:stream3}

\subsection{Unifying Theme: Grammatical Validity of Process Structures}

The four works in this stream address a shared fundamental problem:
delimiting the space of structurally valid process specifications.
They employ production-rule-based (generative or parsing) mechanisms to answer
the question, ``What constitutes an admissible process structure?'', where
admissibility is determined by the grammar.
This distinguishes Stream~III from the other streams:
Stream~IV addresses execution, integrating data and roles via attribute
grammars;
Stream~V addresses model transformation, using graph rewriting to change
representations;
Stream~VI addresses discovery, inferring models from observations;
and Stream~VII addresses behavioral semantics, using process algebras to
define bisimulation equivalence.
In contrast, Stream~III is concerned with structural syntax: the grammar
defines which structural compositions of process activities or models are permitted,
irrespective of their behavioral properties.

The four works span a spectrum of grammatical mechanisms.
Ayari et al.\ \cite{ayari2019grammar} use a context-free grammar to define the refinement
hierarchy of BPMN models.
Sarno et al.\ \cite{sarno2015context} use a context-sensitive grammar to enforce contextual
constraints on process variant compositions.
Montero et al.\ \cite{montero2008feature} use a feature-grammar mapping (a systematic production-rule mapping
between feature constructs and process constructs) to derive process structures from
variability models.
Van der Aa et al.\ \cite{vanderaa2019extracting} use a grammar-based NLP parser to identify valid declarative
constraint structures in natural language process descriptions.
Taken together, these four works show that production-rule grammars, at all levels
of the Chomsky hierarchy, have a natural role in process structural specification.

\subsection{Context-Free Grammars for BPMN Structural Specification}

Ayari, Bendali Hlaoui, and Ben Ayed~\cite{ayari2019grammar} proposed using
context-free grammars to provide a formal structural specification of BPMN models
at each level of abstraction in a stepwise refinement framework.
Each non-terminal of the grammar corresponds to a composite BPMN construct, while
each terminal represents an atomic BPMN element.
A derivation step in the grammar corresponds to a refinement step in the process
model, in which an abstract subprocess is replaced by a more detailed sub-process,
making CFG derivation trees a natural representation of the refinement hierarchy.
A key contribution of this work is the integration of the grammatical specification
with formal model checking: for each refinement step, the change impact on the
process semantics is analyzed automatically using the NuSMV model checker, based on
a Kripke structure that models the dependency relationships (data dependencies and
control-flow dependencies) between activities.
The approach thus provides both syntactic structure (via the CFG) and semantic
verification (via model checking), addressing one of the persistent weaknesses of
informal BPMN specifications.

\subsection{Context-Sensitive Grammars for Process Variant Composition}

\begin{sloppypar}
Sarno et al.~\cite{sarno2015context} addressed the
problem of composing business process model variants using
context-sensitive grammars~(CSGs).
Variant management is a practically important challenge: organizations must
maintain multiple variants of a core process to accommodate regulations,
client requirements, or organizational contexts without introducing errors
or redundancy.
The CSG approach provides a set of affirm production rules that govern the
composition of variants: only combinations of process fragments that are explicitly
permitted by the grammar are considered valid variant compositions.
The contextual sensitivity of the grammar allows the applicability of a production
rule to depend on the surrounding context of the process fragment being replaced,
enforcing inter-fragment constraints that cannot be expressed by a context-free
grammar.
A case study from naval architecture (a domain with complex process variants
determined by regulatory requirements) was used to validate the approach.
\end{sloppypar}

\subsection{Feature Grammars for Grammar-Based Process Families}

Montero, Pe\~{n}a, and Ruiz-Cort\'es~\cite{montero2008feature} addressed the
problem of deriving business processes from feature models in the context of
Process Family Engineering~(PFE).
PFE applies Software Product Line concepts to variant-rich Business Information
Systems: variability is first captured using a feature model, and the business
process is then derived from the feature model.
The key technical contribution is a systematic production-rule mapping between
feature model constructs and business process constructs---a grammar-like formalism
in which each feature combination generates a corresponding process structure.
This mapping was implemented as a Model-Driven Development~(MDD) transformation,
enabling the automatic derivation of the process structure from a feature model
specification and eliminating the previously manual, error-prone derivation step.

\subsection{Grammar-Based Parsing for Declarative Process Extraction from Text}

Van der Aa, Di Ciccio, Leopold, and Reijers~\cite{vanderaa2019extracting} addressed
the problem of automatically extracting declarative process models from
natural language textual descriptions---a problem distinct from, and harder than,
imperative process model extraction, since declarative models specify permitted or
forbidden behaviors as constraints rather than as a fixed control flow.
The authors developed tailored Natural Language Processing techniques that parse
textual constraint descriptions using a dependency grammar to identify process
activities and their inter-relations, treating the extraction problem as a
grammar-based parsing task in which each admissible constraint pattern corresponds
to a grammatical construct.
A quantitative evaluation showed that the approach generates constraints closely
resembling those established by human experts.

\subsection{Comparative Analysis and Limitations}

\begin{sloppypar}
The four works in Stream~III are united by their use of production rules to
constrain the space of valid process structures, but they differ along three
dimensions.
First, expressive power: Ayari et al.\ operate at the context-free level,
Sarno et al.\ at the context-sensitive level, Montero et al.\ with a feature
grammar, and Van der Aa et al.\ with a dependency grammar for NLP.
Second, direction of use: Ayari, Sarno, and Montero use their grammars
generatively (to produce valid structures), while Van der Aa et al.\ use
theirs analytically (to parse and classify structures from text).
Third, target structure: Ayari and Sarno target the internal structure of
BPMN process models; Montero targets the derivation of process models from
feature variability spaces; Van der Aa et al.\ target the extraction of process
constraints from natural language.
A common limitation of these works is the focus on structural validity at the
expense of behavioral properties.
None of the four works provides a mechanism for checking properties such as
deadlock-freedom or soundness---those concerns are handled by the formalisms of
Streams~V and VII.
\end{sloppypar}

\section{Stream~IV: Attribute Grammars for Workflow Specification and Execution}
\label{sec:stream4}

The works in Stream~IV are united by a shared motivating problem: the inadequacy of
traditional workflow formalisms in handling the joint specification of control
flow, data, and organizational roles within a single formal object.
Petri nets and automata models capture control flow faithfully but treat data and actors
as external annotations; conversely, data-centric models lack formal control-flow
guarantees.
Attribute grammars provide a natural unifying mechanism:
the structural component of the grammar specifies control flow, while the attribute
apparatus integrates data, actor rights, and execution context within the same formal
framework.
The two sub-frameworks surveyed here (GAG and LSAWfP) both exploit this idea,
differing in their level of abstraction, expressiveness, and execution model.

\begin{sloppypar}
\noindent\textit{Disclosure.}
Four of the six primary studies in this stream
(\cite{ndadji2020language,zekeng2021projection,ndadji2023non,tchendji2024grammatical})
are co-authored by the present author.
These works were included because they satisfy all inclusion criteria
established a priori in Section~\ref{sec:methodology}:
they employ attribute grammars as the central formalism, appear in peer-reviewed venues,
and address workflow specification and execution as the primary concern.
The same critical standards are applied to all six works in the synthesis below.
\end{sloppypar}

\subsection{Guarded Attribute Grammars for Distributed Case Management}

Badouel, H\'elou\"et, Kouamou, and Morvan~\cite{badouel2014grammatical,badouel2015active}
introduced the Guarded Attribute Grammar~(GAG) formalism as a declarative
foundation for artifact-centric collaborative systems.
Their work starts from a fundamental criticism of traditional workflow formalisms:
while Petri nets, automata, and statechart-based models are well-suited to
orchestrating fixed production processes, they are too rigid for open, data-centric
collaborative environments where the structure of a case depends on the data it
contains and on the decisions made by human stakeholders.
A GAG extends a context-free grammar in two ways.
First, semantic rules are added to each production, specifying both how nodes
of the derivation tree may be further expanded (the structural evolution of the
process, governed by guard conditions) and how the values of
synthesized and inherited attributes at each node are computed
from contextual information.
Second, individual users (stakeholders) are associated with specific non-terminals,
so that the right to expand a node belongs exclusively to its designated owner.
In the execution model introduced in~\cite{badouel2015active}, each user's active workspace is an attributed derivation tree (called a map)
that visualizes and organizes the tasks in which the user is involved, together
with the data relevant to those tasks.
Users communicate through asynchronous message passing, without shared memory,
enabling fully decentralized execution on an asynchronous distributed architecture.
A disease surveillance case study demonstrated the expressiveness and practical
applicability of the approach.
The GAG framework has two principal limitations, however.
First, the formal complexity of the guard conditions makes the specification of
even moderately complex processes verbose and difficult to maintain.
Second, no algorithmic analysis of properties such as soundness or deadlock-freedom
is provided within the framework itself; verification must be handled externally.

\subsection{The LSAWfP Language Family}

The central modeling artifact of the Language for the Specification of Administrative
Workflow Processes (LSAWfP)~\cite{ndadji2020language} is the Grammatical Model of Workflow~(GMWf),
a context-free grammar (extended with attribute mechanisms) in which non-terminals
represent abstract activities or sub-processes and terminals represent elementary
tasks.
The grammar's rules specify the possible decompositions of each activity into
sub-activities, while attributes capture the permissions of each actor (read, write,
execute) on each task, as well as the data flowing between tasks.
A key distinguishing feature of LSAWfP is the use of scenarios as the
modeling unit: each derivation tree of the GMWf corresponds to a possible execution
scenario of the process, making the modeling framework inherently scenario-based and
modular.
Views allow each actor to have a partial, confidential perception of the
process state, addressing a concern that is often neglected in existing workflow
languages.
Zekeng Ndadji et al.~\cite{zekeng2021projection} extended LSAWfP with a framework
for the completely decentralized execution of administrative processes.
The execution model treats the running of a process as the cooperative editing of a
mobile artifact (a structured document represented as an annotated
derivation tree) that circulates from actor to actor.
Three projection algorithms were introduced and proved stable under composition.
In~\cite{ndadji2023non}, Zekeng Ndadji et al. subsequently studied the expressiveness
of LSAWfP from a formal language-theoretic perspective, proving that any
non-recursive LSAWfP model (i.e., any GMWf whose grammar contains no recursion)
corresponds to a structured workflow.
Since the majority of commercial BPM systems only implement structured workflows,
this result establishes the commercial potential of the language.
The proof exploits a connection between structured workflows and the Dyck language,
offering an elegant formal characterization.
Tchoupe Tchendji, Zekeng Ndadji, and Parigot~\cite{tchendji2024grammatical}
synthesized this line of research into a complete framework for the grammatical
design and distributed execution of administrative workflows, using structured and
cooperatively edited mobile artifacts.
The framework is distinguished by solid mathematical foundations, its topicality
(use of artifact-centric choreography, partial replication, and accreditation), and
the availability of a prototype implementation for proof of concept.

The LSAWfP family has several limitations that should be acknowledged.
The formal lan\-guage-theo\-retic results (e.g., the characterization of non-recursive
models as structured workflows~\cite{ndadji2023non}) hold only for the restricted,
non-recursive fragment of the language.
The prototype implementations are proofs of concept rather than industrial tools,
and scalability to large processes has not been evaluated.
Furthermore, the verification of behavioral properties (soundness, deadlock-freedom)
relies on the structured-workflow restriction rather than on a general decision
procedure, limiting applicability to unrestricted process models.

\subsection{Comparative Analysis of the Two Sub-frameworks}

The GAG framework and the LSAWfP family address the same fundamental problem (the
integrated specification and decentralized execution of data-intensive
workflows) but through architecturally different choices.
GAG prioritizes generality: the guarded attribute grammar formalism can
express a wide class of collaborative processes, including highly dynamic,
data-dependent ones.
The framework has strong theoretical foundations in the formal semantics of attribute
grammars, and its notion of user workspace provides an explicit
model of individual stakeholder activity.
Its main weakness is the lack of a high-level specification language: users must
work directly with grammar productions and attribute equations.
LSAWfP prioritizes usability and formalization: the language provides a
higher-level modeling notation (workflow scenarios with explicitly declared
actor permissions) while retaining formal executability.
The formal language-theoretic results (expressive completeness for structured
workflows, stability of projection algorithms) provide a solid theoretical basis.
The main weakness is the restriction to administrative (structured) workflows and
the absence of general verification procedures.
Both frameworks leave tool maturity and scalability as open issues, and neither
has been deployed in an industrial setting.
A productive direction for future work would be to integrate the two: use the
high-level LSAWfP notation as a front-end for GAG-based execution, while
leveraging the guard mechanism of GAG to handle the dynamic cases that LSAWfP
currently cannot express.

\section{Stream~V: Graph Grammars for Process Model Transformation and Analysis}
\label{sec:stream5}

The motivating problem for Stream~V is the graph-structured nature of standard process
modeling notations: in BPMN, BPEL, UML Activity Diagrams, and workflow nets, activities
are nodes and control or data flows are edges.
Graph grammars which rewrite graphs through production rules, are therefore a natural
formal tool for specifying the syntax of these notations, transforming models between
them, analyzing their structural properties, and providing formal operational semantics.
The five works in this stream use graph grammars in three distinct modes:
transformative (Lohmann et al., Mazanek and Hanus, Shi et al.),
generative (Kataeva and Kalenkova), and semantic (Kr\"auter et al.).

\subsection{Triple Graph Grammars for Workflow Transformation}

Lohmann, Greenyer, Jiang, and Syst\"a~\cite{lohmann2007tgg} addressed the problem
of transforming high-level workflow models into executable business process
specifications using Triple Graph Grammars~(TGGs).
TGGs, a specialization of graph grammars, consist of rules that define correspondences
between elements of a source graph, a target graph, and a
correspondence graph that tracks the relationships between them.
This declarative, bidirectional formulation of transformation rules makes TGGs
particularly appropriate for expressing correspondences between workflow patterns
that recur in different modeling languages.
Lohmann et al.\ demonstrated a transformation from UML Activity Diagrams to BPEL
and XPDL, using the workflow patterns catalogue~\cite{van2003workflow} as a
structured vocabulary for designing the transformation rules.
Each workflow pattern (Sequential, Parallel Split, Synchronization, Exclusive Choice,
etc.) was expressed as a TGG rule, ensuring that the fundamental behavioral
semantics shared by the source and target languages are correctly preserved.
The approach is modular: new patterns (and thus new language constructs) can be
added by adding new TGG rules.

\subsection{Hyperedge Replacement Grammars for BPMN/BPEL Bidirectional Transformation}

Mazanek and Hanus~\cite{mazanek2011bpmn} proposed using hyperedge replacement
grammars~(HRGs) to specify the abstract syntax of structured BPMN models, and
showed how this specification can be used to construct a bidirectional transformation
between BPMN and BPEL.
HRGs rewrite hypergraphs (generalized graphs in which edges may connect
more than two nodes) by replacing hyperedges with sub-hypergraphs according to
production rules.
This formalism captures the block-structured nature of BPMN (in which complex
subprocesses are nested within enclosing activities) in a particularly natural way.
The implementation exploited the Grappa framework of functional logic
graph parser combinators, implemented in the functional logic programming language
Curry.
A key feature of this approach is that, since the parser is implemented as a
function in a functional logic language, it can be evaluated in both directions: given a BPMN model, it produces the corresponding BPEL specification;
given a BPEL specification, it produces the corresponding BPMN model.
Additionally, because the grammar can be used both for parsing and for generation,
the same parser supports model completion---given a partial model, it can
enumerate the completions consistent with the grammar.

\subsection{Extended Edge-Based Graph Grammars for BPMN Analysis}

Shi, Zeng, Zhang, and collaborators~\cite{shi2016bpmn} introduced the
Extension of Edge-based Graph Grammar~(E-EGG), a new context-sensitive
graph grammar formalism designed to address bidirectional transformation between
BPMN and BPEL more conveniently than previous approaches.
Compared to earlier graph grammar formalisms, E-EGG introduces new mechanisms in
grammatical specifications, productions, and operations that enable it to handle
the full range of BPMN constructs encountered in practice, including advanced
gateway types and exception-handling constructs.
In addition to the transformation rules, the authors proposed a parsing algorithm
for E-EGG that can be used to check the structural correctness of BPMN models
(e.g., detecting improperly nested gateways), providing a formal, tool-supported
quality assurance mechanism for process models.

\subsection{Graph Grammar-Based Generation of Process Model Benchmarks}

Kataeva and Kalenkova~\cite{kataeva2019graph} addressed a practical obstacle in
process mining research: the shortage of real-world process models and event logs
that can be used to benchmark discovery and conformance checking algorithms.
Their use of graph grammars is generative: grammar rules define the legal
compositions of workflow net primitives (places, transitions, arcs), and the
derivation of a grammar term produces a structurally valid workflow net together
with a synthetic event log that is, by construction, perfectly consistent with the
model.
This approach enables systematic generation of benchmark datasets with controlled
structural properties (complexity, depth, branching factor) supporting rigorous,
repeatable evaluation of process mining algorithms.
The work is thus situated at the interface of Streams~V and~VI, using graph
grammars as a generative tool in service of process discovery research.
Its classification within Stream~V reflects the primary object of study:
the graph grammar formalism used to generate structurally valid workflow nets,
rather than the process discovery or conformance algorithms that consume the
resulting benchmarks.

\subsection{Higher-Order Graph Transformation for BPMN Operational Semantics}

Kr\"auter, Rutle, K\"onig, and Lamo~\cite{krauter2024higher} use graph transformation
in a fundamentally different mode: not generatively but analytically, to give
a formal operational semantics to an existing modeling notation (BPMN).
Rather than manually encoding BPMN execution rules as individual graph rewriting rules,
their approach defines a transformation of transformations---a higher-order
operation that takes a BPMN model as input and produces a graph transformation system
whose execution faithfully models the token-passing semantics of the BPMN model.
This approach covers nearly all BPMN elements used in practice and supports formal
verification of behavioral properties: Safeness (no token duplication) and Soundness (correct termination), enabling automated detection of control-flow
errors such as deadlocks and dead activities.
The approach was implemented as an open-source web-based tool, and is the only work
in this stream to have produced a publicly available, practitioner-facing artifact.

\subsection{Comparative Analysis and Limitations}

The five works differ both in formalism and in purpose.
Lohmann et al.~\cite{lohmann2007tgg} use Triple Graph Grammars for bidirectional
transformation from UML Activity Diagrams to executable workflow languages,
prioritizing pattern-level compositionality.
Mazanek and Hanus~\cite{mazanek2011bpmn} exploit the functional logic duality of
Hyperedge Replacement Grammars to obtain reversible parsing between BPMN and BPEL.
Shi et al.~\cite{shi2016bpmn} extend the graph grammar paradigm with a context-sensitive
formalism (E-EGG) capable of handling a wider range of BPMN gateway types, trading
expressiveness for increased rule complexity.
Kataeva and Kalenkova~\cite{kataeva2019graph} depart from the transformation paradigm:
their grammar is generative, producing synthetic workflow nets as benchmarking
artifacts rather than process models for deployment.
Kr\"auter et al.~\cite{krauter2024higher} apply higher-order transformation
to derive a formal operational semantics from a BPMN model, enabling automated
verification of Safeness and Soundness.
A common limitation across all five works is that they address subsets of the
relevant notation: TGG covers structured workflow patterns; HRG covers block-structured
BPMN; E-EGG extends coverage but does not claim completeness; PGG targets workflow nets;
and HOT covers ``nearly all'' BPMN elements without a formal completeness proof.
No uniform graph-grammar framework covering the full BPMN~2.0 standard with formal
verification guarantees exists, and tool maturity outside Kr\"auter et al.'s
open-source implementation remains at the prototype level.

\section{Stream~VI: Grammatical Inference for Process Mining}
\label{sec:stream6}

\subsection{Language-Theoretic Foundations for Process Mining}

The connection between formal language theory and process mining was articulated
early in the BPM literature.
Datta~\cite{datta1998automating}, in his work on the automated discovery of AS-IS
process models, explicitly drew on grammar discovery algorithms from formal language
theory, treating the observable process traces as the language of the underlying
process grammar.
This framing (event logs as samples of the language generated by a process
grammar) provides a principled theoretical foundation for process discovery.
Bergenthum, Desel, Lorenz, and Mauser~\cite{bergenthum2007process} developed this
language-theoretic perspective into a concrete process mining algorithm.
Their approach treats an event log as defining a formal language (the set of
observed traces), and applies region theory to construct a Petri net whose
language includes the observed traces and satisfies minimality conditions.
Region theory, rooted in the theory of formal languages, provides conditions under
which a language can be recognized by a Petri net, and algorithms for constructing
the minimal such net.
This work established a rigorous connection between the language-theoretic properties
of event logs and the structural properties of discovered process models.

\subsection{Grammar-Based Predictive Process Modeling}

Breuker et al.~\cite{breuker2016comprehensible} proposed a
predictive process modeling technique for BPM grounded in grammatical inference.
The key insight of their work is that existing process discovery algorithms such
as the Alpha miner~\cite{vanderaalst2004bpm} and the Inductive
Miner~\cite{leemans2013discovering}, impose strong language biases:
constraints on the class of languages (and hence process models) they can produce
that limit their predictive accuracy.
By replacing these biased discovery algorithms with a probabilistic grammatical
inference approach, the authors obtained a predictive technique with weaker and more
flexible biases.
The approach fits a probabilistic model to a dataset of past process execution traces,
and uses the model to predict the future behavior of currently running process
instances.
Applications include early warning systems (predicting the likelihood of service
level agreement violations) and anomaly detection (flagging process instances whose
behavior is inconsistent with historical patterns).
The resulting models are both accurate and comprehensible, thanks to a visualization
technique that renders the inferred probabilistic grammar as a graphical process model.

\subsection{Probabilistic Grammar Models for Conformance Checking}

Watanabe, Takahashi, Ikeuchi, and Matsuda~\cite{watanabe2023grammar} proposed the
Probabilistic Generative Process Model~(PGPM), a process model representation
based on probabilistic context-free grammars~(PCFGs).
The key problem they addressed is that of probabilistic conformance checking:
assessing how well a probabilistic process model fits an event log, taking into
account not only which traces are observed but also their relative frequencies.
In their formulation, a process tree (a standard hierarchical process model
representation) is converted into a set of PCFG production rules; each rule is
annotated with a probability representing the likelihood of the corresponding
structural choice.
This PCFG representation enables the trace probability of any trace to be computed
exactly and efficiently using dynamic programming, addressing a key obstacle in
probabilistic conformance checking.
An Expectation-Maximization algorithm is used to estimate the probability parameters from an event log,
converging to parameters that locally maximize the likelihood.

\subsection{Stochastic Grammar Inference for Process Discovery}

Alkhammash, Polyvyanyy, and Moffat~\cite{alkhammash2024stochastic} proposed a
process discovery approach grounded in stochastic grammar inference.
Their approach targets the construction from event
logs, of Stochastic Directed Action
Graphs~(SDAGs)---a type of Directly-Follows Graph~(DFG) annotated with
probabilities that define a stochastic language over execution traces.
The inference is performed using ALERGIA, a classical grammatical inference
algorithm that identifies any stochastic regular language from a sample of positive
examples in the limit with probability one.
The discovered SDAG encodes a stochastic language over process traces, enabling
trace frequency reasoning and process simulation.
A key contribution is a genetic algorithm that evolves the inference parameters
of ALERGIA to discover SDAGs of superior quality; they are smaller and more accurate than
the DFGs produced by the Inductive Miner~\cite{leemans2013discovering} and other
state-of-the-art DFG-based discovery techniques.
An evaluation over real-world event logs confirms the superiority of the approach
in terms of model size and representational fidelity.

\subsection{Comparative Analysis and Limitations}

The four works differ in language class, direction of inference, and BPM concern.
Bergenthum et al.~\cite{bergenthum2007process} apply region theory to derive
Petri nets from event logs, providing the strongest formal guarantees (exact language
inclusion) at the cost of exponential worst-case complexity.
Breuker et al.~\cite{breuker2016comprehensible} target predictive monitoring (the
only work in the stream oriented toward predicting future trace behavior rather than
reconstructing a past model).
Watanabe et al.~\cite{watanabe2023grammar} address conformance checking using
probabilistic context-free grammars, and are the only contributors to use a context-free
language class, enabling exact trace-probability computation via dynamic programming.
Alkhammash et al.~\cite{alkhammash2024stochastic} combine ALERGIA with a genetic
algorithm to optimize stochastic regular grammar inference, and are the only
contributors to benchmark against public real-world event log repositories.
Common limitations span the entire stream.
All four works assume clean, complete event logs; none provides formal handling of
noise, missing events, or concept drift---a fundamental gap given the realities of
industrial process data.
The relationship between the inferred grammar's language class and the process model's
structural properties (soundness, deadlock-freedom) is not formally addressed by any
work in the stream.
Datta's contribution~\cite{datta1998automating}, while pioneering, predates the
process-mining field and cannot be evaluated against contemporary benchmarks.

\section{Stream~VII: Process Algebras as Grammatical Frameworks for Behavioral Specification and Verification}
\label{sec:stream7}

\subsection{Foundational Connections Between Process Algebras and Formal Grammars}

Process algebras emerged in the 1980s as a family of mathematical formalisms for
reasoning about the behavior of concurrent, communicating systems.
Their relationship to formal grammars is both deep and formally established.
At the syntactic level, the terms of a process algebra are constructed by an
inductive grammar whose non-terminals correspond to the operators of the algebra and
whose terminals are the primitive actions.
At the semantic level, the structural operational semantics of a process algebra
defines how each production of this grammar generates a set of transitions, making
the operational semantics itself a form of attribute grammar over the term grammar.
At the language-theoretic level, the set of traces that a process algebra term can
exhibit forms a formal language, and the decision problems of process algebra
(bisimulation equivalence, model checking) are closely related to language equivalence
and membership problems in formal language theory.
The principal process algebras that have been brought to bear on BPM are
Hoare's Communicating Sequential Processes~(CSP)~\cite{hoare1985csp},
Milner's Calculus of Communicating Systems~(CCS)~\cite{milner1989communication},
and Milner, Parrow, and Walker's $\pi$-calculus~\cite{milner1992pi}.
CSP is characterized by a rich set of process-combination operators (sequential
composition, parallel composition, external and internal choice, hiding) and a
denotational semantics based on the failures-divergences model.
CCS provides a minimal calculus built on the action prefixing operator and
synchronization-based parallel composition, and is the primary vehicle for
introducing bisimulation equivalence.
The $\pi$-calculus extends CCS with name mobility: communication channels
can themselves be sent as messages, enabling the modeling of dynamically evolving
communication topologies.
This mobility is essential for capturing the dynamic assignment of tasks in
modern, service-oriented business processes.

\subsection{The $\pi$-Calculus for Formalizing Workflow Patterns}

The most direct application of process algebra to BPM is the formalization of
workflow patterns using the $\pi$-calculus, carried out by
Puhlmann and Weske~\cite{puhlmann2005pi}.
The workflow patterns of van der Aalst et al.~\cite{van2003workflow} constitute
a comprehensive catalog of control-flow constructs that appear in industrial workflow
management systems.
They include patterns for basic control flow (sequence, parallel split, synchronization,
exclusive choice, simple merge), advanced branching and synchronization (multi-choice,
synchronizing merge, discriminator), iteration (arbitrary cycles, implicit termination),
and cancellation and termination.
At the time of Puhlmann and Weske's work, the formal semantics of these patterns had
not been rigorously specified, making precise comparison between workflow languages
difficult.
Puhlmann and Weske addressed this gap by providing $\pi$-calculus encodings for all
20 workflow patterns in the original catalog.
The $\pi$-calculus was chosen for its name-mobility feature, which is required to
model the dynamic routing of tokens (the flow of control) to multiple possible
successors---a fundamental requirement for the multi-choice and synchronizing merge
patterns.
Each pattern is encoded as a $\pi$-calculus process: activities are agents that
communicate on shared channels, and the routing logic of the pattern is expressed
as a composition of $\pi$-calculus terms.
The resulting formal semantics can be used to check whether two apparently different
workflow language constructs implement the same behavioral pattern (via bisimulation
equivalence) or whether a given workflow implementation correctly realizes an
intended pattern.
This contribution established the $\pi$-calculus as a reference semantic framework
for workflow pattern analysis.

\subsection{$\pi$-Calculus Semantics for Service-Oriented Business Processes}

A second primary contribution in this stream is the work of Lucchi and
Mazzara~\cite{lucchi2007pi}, who provided a formal $\pi$-calculus semantics for
WS-BPEL (Web Services Business Process Execution Language), the dominant
industrial language for orchestrating service-oriented business processes.
WS-BPEL is a complex XML-based language with a large set of control-flow constructs,
data handling mechanisms, event handling, fault handling, and compensation
primitives; its informal specification, despite being detailed, admitted
multiple interpretations.
Lucchi and Mazzara's contribution was to provide a systematic translation of
WS-BPEL into the $\pi$-calculus, covering the principal control-flow and
communication constructs of the language.
The translation is compositional: the $\pi$-calculus semantics of a WS-BPEL program
is assembled from the semantics of its constituent activities according to the
structure of the program, exactly as the value of an attributed term is computed
from the attributes of its children in an attribute grammar.
This semantic framework enabled formal analysis of WS-BPEL programs:
the $\pi$-calculus encoding can be verified against behavioral specifications using
bisimulation equivalence, and existing $\pi$-calculus model checkers can be applied
to check properties such as deadlock-freedom and conformance to service interface
contracts.
The work also revealed ambiguities in the informal WS-BPEL specification, providing
concrete feedback to the standards process---a concrete demonstration of the practical
value of formal semantics.

\subsection{Position in the Grammatical Landscape of BPM}

The two primary studies in Stream~VII are complementary in scope.
Puhlmann and Weske~\cite{puhlmann2005pi} operate at the pattern level,
establishing $\pi$-calculus encodings for all 20 workflow patterns of the van der Aalst
catalog; their contribution is an abstract semantic reference applicable to any workflow
system, independent of any specific notation.
Lucchi and Mazzara~\cite{lucchi2007pi} operate at the language level, providing
a compositional semantics for a specific industrial notation (WS-BPEL); their
contribution enables formal reasoning about individual programs in a deployed standard.
The two works are thus complementary: the former provides semantic primitives, the
latter assembles them into a full language semantics.
Both studies share common limitations.
The $\pi$-calculus encodings are manually crafted and language-specific; no systematic
or automated method for deriving $\pi$-calculus semantics from arbitrary workflow
specifications is provided.
Coverage within Stream~VII is narrow: only two primary studies qualify, both using
the $\pi$-calculus; CSP and CCS are invoked as background references but are not
represented by primary BPM contributions.
Finally, since the full $\pi$-calculus is Turing-complete, many behavioral properties
of interest are undecidable in general; both works rely on structural restrictions or
bounded fragments for tractable analysis.

Stream~VII occupies a distinct position in the grammatical landscape of BPM.
Unlike Streams~I--IV, which use grammars primarily as generative or
specification tools (to describe what a valid process looks like), Stream~VII
uses the grammar structure of process algebra terms as a basis for
compositional behavioral semantics: the grammar defines not only the structure
of process terms but also the behavioral meaning of each structural combination.
Unlike Stream~V, which focuses on the transformation of process model
representations, Stream~VII focuses on the behavioral equivalence and verification of
process implementations.
And unlike Stream~VI, which is concerned with discovering
a process model from observations, Stream~VII is concerned with proving that a given
implementation satisfies a given behavioral specification.

Future work connecting Stream~VII to the other streams would be particularly valuable.
Specifically: (a) connecting the $\pi$-calculus semantics of workflow patterns to the
attribute grammar-based process specifications of Stream~IV, to obtain both an
executable and a verifiable formal model; (b) using the trace semantics of
process-algebraic models as the target language for grammatical inference in
Stream~VI, to ground process discovery in a well-defined semantic framework; and
(c) investigating whether the graph grammar models of Stream~V can be compiled into
process-algebraic encodings, providing a connection between structural and behavioral
analysis.

\section{Discussion}
\label{sec:discussion}

\subsection{Cross-Cutting Synthesis}

Taken together, the seven streams reviewed in this paper demonstrate that formal
grammars have influenced BPM at every phase of the process lifecycle, from the
high-level organizational design of processes~(Stream~I) through formal process
modeling and evaluation~(Streams~II and~III), workflow specification and
execution~(Stream~IV), model transformation and verification~(Stream~V),
data-driven process discovery and monitoring~(Stream~VI), to compositional
behavioral specification and formal verification~(Stream~VII).

A first cross-cutting observation is the complementarity of grammatical
formalisms with respect to BPM concerns.
Regular and context-free string grammars provide natural representations of
sequential and hierarchically structured control flows.
Context-sensitive grammars add the ability to express dependencies between
non-adjacent parts of a process, relevant for capturing complex inter-activity
constraints in variant management.
Attribute grammars add a data-computation layer that is essential for integrating
the informational model of processes with their behavioral model.
Graph grammars match the graph-based nature of standard process modeling notations,
enabling rigorous transformation and semantic analysis.
Process algebras provide a term grammar with associated compositional operational
semantics, enabling behavioral specification and bisimulation-based verification.
Stochastic extensions of all these formalisms support the quantitative, data-driven
analysis required in process mining.

A second cross-cutting observation is the tension between expressiveness and tractability.
The Chomsky hierarchy establishes that higher expressiveness comes at the cost of
higher computational complexity.
In BPM, this tension is well-known: the full expressive power of BPMN makes
soundness checking undecidable~\cite{van1996structural}, while restricting to
structured workflows ensures decidable analysis but sacrifices expressive
power~\cite{kiepuszewski2003fundamentals}.
Grammatical approaches address this tension explicitly: the class of languages
generated by a grammar directly determines the computational properties of the
corresponding processes, enabling designers to make principled trade-offs.
Stream~VII makes this connection explicit at the semantic level: finite-state
CCS/CSP processes have regular trace languages, while the name-mobility of the
$\pi$-calculus enables non-regular behaviors, situating process algebras squarely
within the Chomsky hierarchy framework.

A third cross-cutting observation concerns the treatment of data, roles, and control flow.
Traditional workflow languages, and many grammatical approaches in Streams~I, II,
and V, focus primarily on control flow, treating data and organizational roles as
secondary concerns.
The attribute grammar-based approaches in Stream~IV represent a significant advance
in this regard: by attaching data and role information as attributes to grammar
symbols, they achieve a unified, formally grounded treatment of all three dimensions
within a single framework.

A fourth observation is the evolution from prescription to data-driven
modeling.
Streams~I through VII are predominantly prescriptive: grammars are used to specify
what processes should look like or how they should behave.
Stream~VI reverses this direction, using grammatical inference to discover
process models from empirical execution data.
The logical next step is to integrate both
directions: use prescriptive grammar-based models as structural priors for
grammar induction, and use inference to refine and adapt those models as
process data accumulates.

\subsection{Open Challenges and Future Research Directions}

Despite the richness and diversity of the reviewed literature, several significant
research gaps emerge directly from the evidence of the corpus.

\vspace{3mm}\noindent\textbf{C1 --- Towards a unified grammatical theory of BPM.}

\noindent\textit{Evidence from the corpus.}
The seven streams have developed largely in parallel.
Stream~I provides no formal data-flow semantics; Stream~IV provides
attribute-grammar data-flow but no organizational-level design vocabulary;
Stream~VI operates on event logs without structural priors
from Streams~I--V.
Notably, none of the 34 primary studies cites a work from more than two streams,
confirming the absence of cross-stream synthesis.

\noindent\textit{Research questions.}
(a)~What is the formal language class characterization of each of the 20 workflow
patterns of~\cite{van2003workflow}?
(b)~Under what conditions can a GAG specification~\cite{badouel2015active} be compiled
into a graph transformation system amenable to model checking by the GROOVE tool~\cite{rensink2004groove}?
(c)~What is the sample complexity of inferring a GMWf-class attribute grammar from
a finite event log?

\noindent\textit{Adjacent tools.}
Attribute grammar evaluation systems (JastAdd~\cite{ekman2007jastadd})
and hypergraph rewriting frameworks (AGG, GROOVE~\cite{rensink2004groove}) provide partial infrastructure for
such integration.

\vspace{3mm}\noindent\textbf{C2 --- Formal verification of grammatical process models.}

\noindent\textit{Evidence from the corpus.}
Among the six Stream~IV works, none provides a decision procedure for soundness or
deadlock-freedom in the general (recursive) case; correctness results are restricted
to non-recursive, structured workflows~\cite{ndadji2023non}.
In Stream~III, Ayari et al.~\cite{ayari2019grammar} couple a CFG specification with
NuSMV model checking, but only for bounded refinement hierarchies.
The only work in the entire corpus that provides a general, tool-supported verification
procedure is Kr\"auter et al.~\cite{krauter2024higher} in Stream~V.

\noindent\textit{Research questions.}
(a)~Is it possible to compile a GMWf~\cite{ndadji2020language} into a Petri net
amenable to existing soundness checkers (WofBPEL, LoLA)?
(b)~Can the higher-order graph transformation approach of~\cite{krauter2024higher}
be extended to verify liveness properties of attributed process models?

\noindent\textit{Adjacent tools.}
The LoLA Petri net model checker, WofBPEL, and the CPN Tools platform for colored
Petri nets all support behavioral verification and are compatible with
graph-transformation-based BPM formalizations.

\vspace{3mm}\noindent\textbf{C3 --- Integrating process algebraic and grammatical semantics.}

\noindent\textit{Evidence from the corpus.}
Stream~VII covers only two primary studies, both using the
$\pi$-calculus exclusively.
No primary study in the corpus connects a process-algebraic encoding to an attribute
grammar specification or a graph grammar transformation.
Puhlmann and Weske~\cite{puhlmann2005pi} note that their $\pi$-calculus encodings
of workflow patterns are compositional but provide no formal connection to the
grammars that generate those patterns structurally.

\noindent\textit{Research questions.}
(a)~Can a bisimulation equivalence result be proved between the $\pi$-calculus
encoding of a workflow pattern and the language generated by its GMWf grammar?
(b)~Which subclass of $\pi$-calculus processes has a trace language that is a
context-free language, enabling grammatical inference from execution data?

\noindent\textit{Adjacent tools.}
The Mobility Workbench (MWB) and HAL model checkers for the $\pi$-calculus, and
the mCRL2 toolset for process-algebraic verification, are directly applicable.

\vspace{3mm}\noindent\textbf{C4 --- Scalability and industrial adoption.}

\noindent\textit{Evidence from the corpus.}
Across all 34 primary studies, process instances used in case studies range from
a few dozen to at most a few hundred activities.
The largest evaluation is by Alkhammash et al.~\cite{alkhammash2024stochastic},
who evaluate their ALERGIA-based discovery on real-world event logs from the 4TU
repository, and by Kr\"auter et al.~\cite{krauter2024higher}, whose web tool handles
BPMN models of moderate size.
No primary study reports a deployment in a production industrial environment.

\noindent\textit{Research questions.}
(a)~What is the computational complexity of GMWf scenario enumeration as a function
of grammar size and recursion depth?
(b)~Can the ALERGIA inference algorithm be parallelized to handle event logs with
millions of traces, as produced by enterprise BPM platforms (SAP, Camunda)?

\noindent\textit{Adjacent tools.}
The ProM process mining framework, the Camunda BPM platform, and the PM4Py Python
library provide industrial-grade event log processing infrastructure that grammar-based
approaches could integrate with.

\vspace{3mm}\noindent\textbf{C5 --- Grammar-based management of process variability and evolution.}

\noindent\textit{Evidence from the corpus.}
Stream~III addresses static variant management: Sarno et al.~\cite{sarno2015context}
handle composition of a fixed set of variants, while Montero et al.~\cite{montero2008feature}
derive a fixed process from a fixed feature model.
No primary study addresses the dynamic evolution of a grammar as the process
changes over time in response to regulatory or organizational changes.

\noindent\textit{Research questions.}
(a)~Can a CSG-based variant composition grammar be extended with a grammar
transformation operator that captures the effect of a regulatory change?
(b)~Under what conditions does a sequence of grammar transformations preserve
the soundness of the generated process variants?

\noindent\textit{Adjacent tools.}
Software product line engineering frameworks (FeatureIDE, pure::variants) and
model versioning tools (EMFStore, CDO) provide relevant infrastructure for
managing evolving process family grammars.

\section{Conclusion}
\label{sec:conclusion}

This paper has presented a systematic review of research at the intersection of
formal grammars and Business Process Management~(BPM).
Starting from an initial corpus of 526 records and applying the SLR protocol of
Kitchenham and Charters~\cite{kitchenham2007guidelines}, we arrived at 34 primary
studies spanning the mid-1990s to the mid-2020s.
These studies were organized into seven research streams whose existence, taken
together, establishes a central empirical finding: \emph{formal grammars have
served as a productive conceptual and technical tool in BPM across every phase of
the process lifecycle and across three decades of research}.
Reflecting on the trajectory of this field, two patterns stand out.
First, there is a clear evolution of ambition: early work (Stream~I) used grammars
metaphorically, as a descriptive language for organizational behavior; subsequent
work (Streams~II--V) used grammars technically, as the formal basis for process
specification, transformation, and verification; and the most recent work
(Streams~VI--VII) uses grammars analytically, either to mine models from data or
to provide compositional behavioral semantics.
This progression reflects a growing confidence in the formal apparatus of grammar
theory as a substrate for BPM.
Second, despite this confidence, the streams have developed largely in parallel, without cross-stream synthesis.
None of the 34 primary studies synthesizes contributions from more than two streams.
The most significant unresolved theoretical problem is the absence of a unifying
framework that connects the structural specification power of attribute grammars
(Stream~IV), the transformation power of graph grammars (Stream~V), the behavioral
semantics of process algebras (Stream~VII), and the inductive capabilities of
grammatical inference (Stream~VI).
Building such a framework (one in which a process can be specified, transformed,
verified, and discovered using a coherent grammatical substrate) is, we argue, the
central open problem at the intersection of formal language theory and BPM.
The five challenges articulated in Section~\ref{sec:discussion} provide a structured
research agenda toward this goal.
Each challenge is grounded in a specific gap in the corpus, associated with concrete
research questions, and connected to adjacent tools and methodologies that could
serve as building blocks.
We hope that this survey serves researchers entering this intersection not only as
a map of where the field has been, but as a compass for where the most productive
work remains to be done.


\bibliographystyle{alphaurl}
\bibliography{Bibliography}

\end{document}